\documentclass{PoS}

\title{Exotic hadron production in hard exclusive reactions}

\ShortTitle{Exotic hadron production in hard exclusive reactions}

\author{\speaker{Hiroyuki Kawamura}
       \thanks{H.K. is supported by a Grant-in-Aid for Scientific Research
               on Priority Areas ``Elucidation of New Hadron with a Variety of 
               Flavors (E01:21105006).''}\\
        KEK Theory Center, Institute of Particle and Nuclear Studies, 
          KEK, 1-1, OHO, Tsukuba, Ibaraki, 305-0801, Japan\\
        E-mail: \email{hiroyuki.kawamura@kek.jp}}

\author{ Shunzo Kumano
         \thanks{S.K. is supported in part by a Grant-in-Aid for Scientific 
         Research (No.25105010).}\\
         KEK Theory Center, Institute of Particle and Nuclear Studies, 
         KEK, 1-1, OHO, Tsukuba, Ibaraki, 305-0801, Japan\\
         J-PARC Branch, KEK Theory Center, Institute of Particle and 
         Nuclear Studies, KEK and Theory Group, 
         Particle and Nuclear Physics Division, J-PARC Center, 203-1, 
         Shirakata, Tokai, Ibaraki, 319-11-6, Japan\\ 
        E-mail: \email{shunzo.kumano@kek.jp}}

\author{ Takayasu Sekihara\\
         KEK Theory Center, Institute of Particle and Nuclear Studies, 
         KEK, 1-1, OHO, Tsukuba, Ibaraki, 305-0801, Japan\\
        E-mail: \email{sekihara@post.kek.jp}}

\abstract{We consider hard exclusive production of exotic hadrons 
to study their internal structure. 
Revisiting the constituent-counting rule for the large-angle exclusive 
scattering, we discuss general features expected for the production 
cross section of exotic hadrons whose leading Fock states are given 
by multi-quark states other than the ordinary baryon ($qqq$) or 
meson ($q\bar{q}$) states.
We take the production of $\Lambda(1405)$ as an example and propose 
to study its partonic configuration from the asymptotic scaling of 
the cross section, which is measurable at J-PARC. 
We also discuss the production of a pair of the light-hadrons such as 
$f_0(980)$s and $a_0(980)$s in $\gamma^*\gamma$ collisions in the framework 
of QCD factorization,  in which the cross section is expressed 
as a convolution of the perturbative coefficients and the generalized 
distribution amplitudes (GDAs). 
We demonstrate how the internal structure of $f_0(980)$ or $a_0(980)$ 
can be explored by measuring the GDAs at $e^+e^-$ experiments such as 
the B-factories.   }

\FullConference{XV International Conference on Hadron Spectroscopy-Hadron 2013\\
		4-8 November 2013\\
		Nara, Japan }

\begin{document}

\section{Introduction}
Despite its non-relativistic nature, the conventional quark model has 
been remarkably successful in classifying the hundreds of the observed 
hadrons into baryons and mesons.
On the other hand, the underlying theory of the strong interaction, 
Quantum Chromodynamics or QCD, does not prohibit hadrons with other 
quark-gluon configurations such as tetraquarks, pentaquarks, glueballs, etc, 
to which we here refer as exotic hadrons. 
Actually, the discovery of ``XYZ" states at the B-factories and other 
facilities in the last decade \cite{belle-babar} has demonstrated 
the existence of exotic hadrons in the heavy-quark sector and has 
motivated extensive theoretical studies to understand their properties 
\cite{Brambilla:2010cs}. 
In the light-quark sector, there have been long-standing candidates for 
exotic hadrons. For example, $\Lambda(1405)$ is hard to understand 
as an excited state of the ordinary $uds$ combination 
since it is much lighter than 
the corresponding excited nucleon $N(1535)$ and the possibility of being 
a $\bar{K}N$ bound state has been discussed \cite{dalitz,jido,sk-2013}. 
Likewise, the light scalar mesons obey an anomalous mass relation 
$M(f_0)\approx M(a_0)>M(\kappa)$ instead of $M(f_0)> M(\kappa)>M(a_0)$ 
which is expected from the conventional quark model, 
and could be understood as tetraquarks \cite{f0-exotic}.

So far, the structure of exotic hadrons have been studied mainly in terms 
of hadronic observables such as mass, spin and decay width. 
Here, we consider the possibility of studying the internal structure 
by hard processes, in which the quarks and gluons are the relevant 
degrees of freedom. 
Since the exotic hadron candidates are unstable particles, we cannot use 
them as a target and need to find out what can imply the "exoticness" 
in the production processes. The inclusive production of exotic hadrons 
was studied in \cite{hkos08}, in which it was demonstrated that 
the signature of a multi-quark configuration of $f_0(980)$ can appear
as the difference between the "favored" and "disfavored" fragmentation function.
For the exclusive hard processes, some studies have already been performed 
for electro- and hadroproduction of $\Theta^+$ pentaquark \cite{diehl-penta}, 
and the production of a hybrid meson in $\gamma^*\gamma$ collisions 
\cite{hybrid}. 
In this work, we explore other possibilities for using exclusive processes 
as a means of studying the partonic structure of exotic hadrons. 
Firstly, we discuss the production of $\Lambda(1405)$ in the large-angle 
exclusive scattering, $\pi+p\rightarrow K+\Lambda(1405)$, in the light of 
the constituent-counting rule \cite{counting-Matveev,counting-Brodsky}. 
Secondly, we discuss the exclusive pair production, 
$\gamma^*+\gamma\rightarrow h+\bar{h}(h=f_0(980)~ $or$~ a_0(980))$, 
in the kinematical region where the process is described in terms of 
the GDAs \cite{muller-1994,diehl-2000}. 
The details are found in \cite{kks-2013,ks-2013}.   

\section{Exclusive production of $\Lambda(1405)$ }
In the framework of QCD factorization \cite{exclusive-Brodsky}, 
the scattering amplitude of the $2\rightarrow 2$ hadronic process 
with a large momentum transfer is expressed schematically as 
\begin{eqnarray}
M(a+b\rightarrow c+d)|_{s,|t|,|u|\gg \Lambda_{QCD}^2}
\approx
\phi_a\times\phi_b\times H(\alpha_s)\times\phi_c\times\phi_d \ , 
\label{large-angle}
\end{eqnarray}
where $s,t,u$ are the Mandelstam variables, $\phi_i$ is a light-cone 
distribution amplitudes (LCDA) of a hadron $i$ and $H(\alpha_s)$ is 
the hard part of the scattering amplitude.   
Although the above expression holds as a general formula, it is difficult 
to predict the absolute value of the cross section in most cases because, 
(i) the LCDAs are not known except for the pion; 
(ii) a large number of Feynman diagrams contribute to the hard part.
Nevertheless, the asymptotic scaling of the cross section can be 
predicted by the constituent-counting rule \cite{counting-Matveev,
counting-Brodsky}:
\begin{eqnarray}
\left.\frac{d\sigma_{ab\rightarrow cd}}{dt}
\right|_{s,|t|,|u|\gg \Lambda_{QCD}^2}
=\frac{1}{s^{n-2}}f_{ab\rightarrow cd}(t/s) ,
\label{counting}
\end{eqnarray}
where $f(t/s)$ is a dimensionless function of the scattering angle, 
and the number $n$ is defined by $n=n_a+n_b+n_c+n_d$ with the $n_i$ being 
the number of constituents in the leading Fock state of the hadron $i$.  
The constituent-counting rule can be understood from the dimensional 
counting of the hard part $H(\alpha_s)$ in eq.(\ref{large-angle}) 
\cite{kks-2013} and
has been confirmed in the experiments at BNL and JLab \cite{bnl-jlab}. 
In addition, we found that the counting-rule holds for the production of 
the ground-state Lambda in $\pi^-+p\rightarrow K+\Lambda$ with 
the scaling factor $n=10.1\pm0.6 $ \cite{kks-2013}. 
In all of the above cases, the transition from the resonance region 
to the scaling region occur at $\sqrt{s}=2-3$GeV.

\begin{figure}
\includegraphics[width=.5\textwidth]{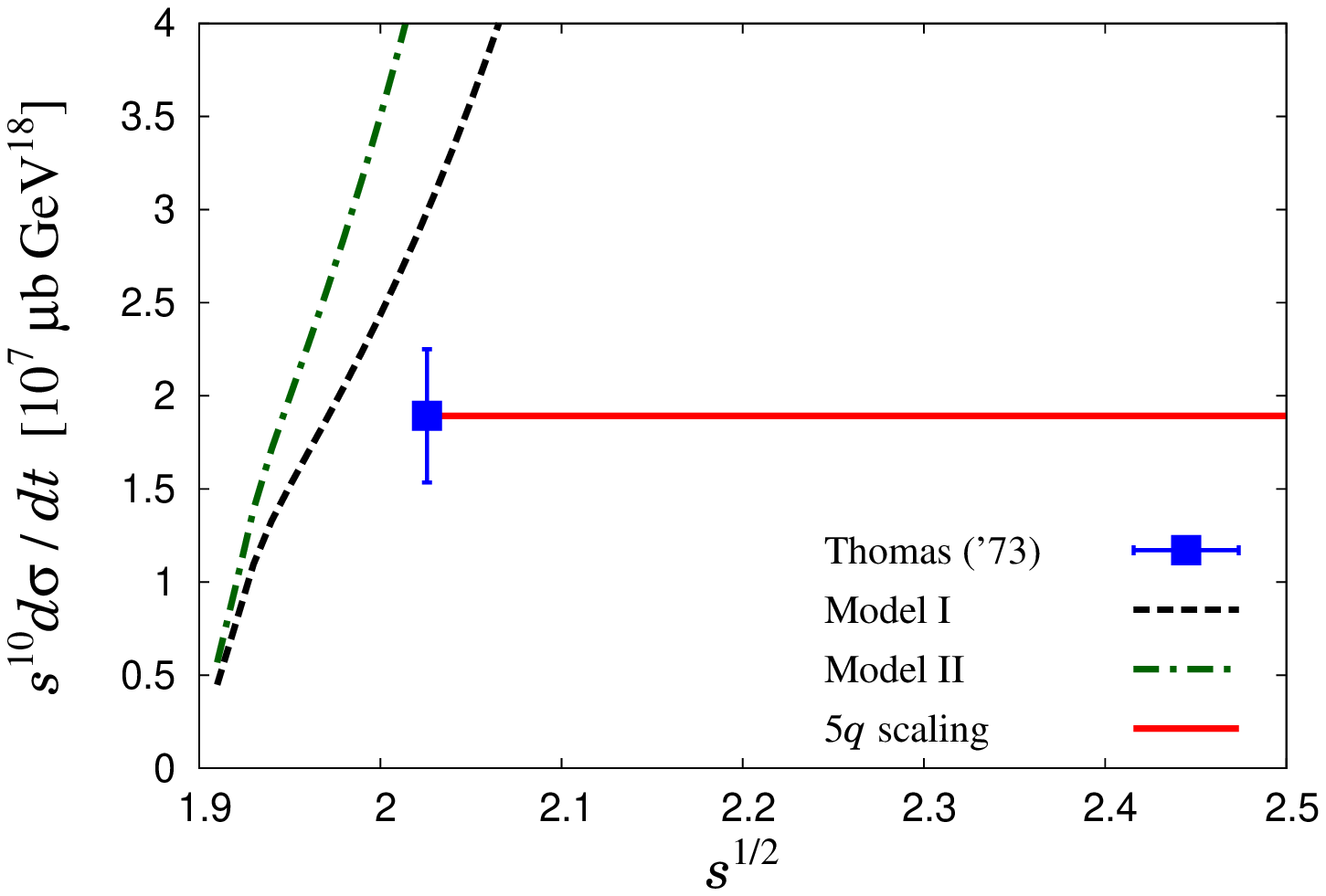}
\includegraphics[width=.5\textwidth]{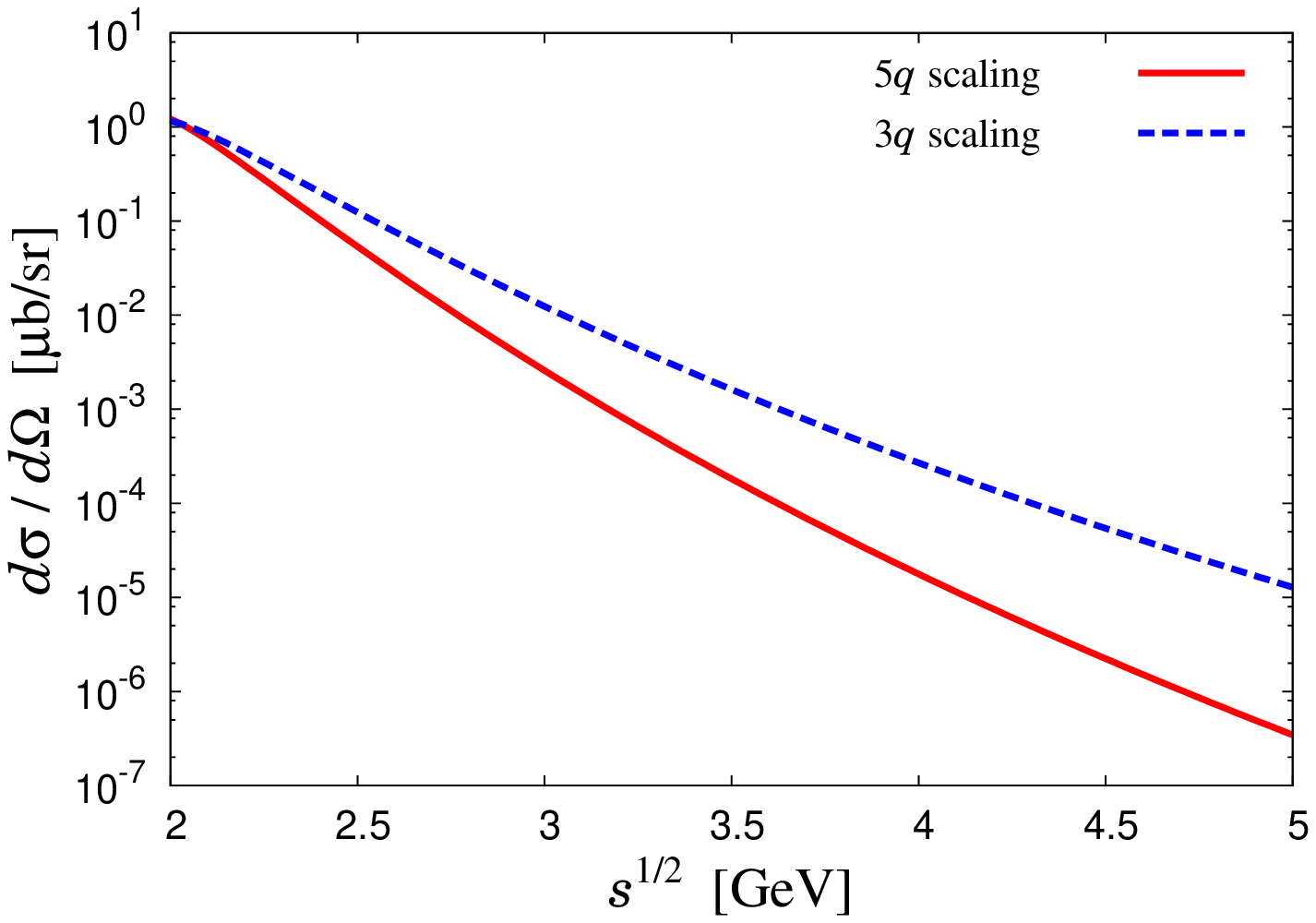}
\caption{Cross section of $\pi^-+p\rightarrow K^0+\Lambda(1405)$ (left) and 
its scaling at high energies (right).}
\label{lambda-1405}
\end{figure}

Figure \ref{lambda-1405} shows the cross section of the exclusive 
$\Lambda(1405)$ production, $\pi^++p\rightarrow K+\Lambda(1405)$.
For this process, there is only one experimental data 
\cite{lambda-1405-prod-ex} which is shown in the left panel of 
Figure \ref{lambda-1405}. The data is roughly consistent with 
the theoretical estimate based on the chiral unitary model 
\cite{lambda-1405-prod-th}.
The solid line in Figure \ref{lambda-1405} is obtained by extrapolating 
the experimental data by assuming 
the leading Fock component of $\Lambda(1405)$ to be a five-quark state, 
so that the total number of constituents is given by $n=2+3+2+5=12$ and 
the scaling rule that $s^{10}d\sigma/dt=$ constant. 
The comparison of the cross sections anticipated when the $\Lambda(1405)$ 
is a three-quark state and a five-quark state is shown in the right panel.
The result indicates that those two cases can be clearly distinguished 
if we have enough data from experiments, and such experiments are 
possible using the high-momentum beam at J-PARC  \cite{J-PARC-exp-chang}.   

\section{Pair production of {\boldmath$f_0(980)$} and 
{\boldmath$a_0(980)$} in the GDA kinematics}
The cross section of the hadron pair production 
$\gamma^*+\gamma\rightarrow h+\bar{h}$ can be described as a production 
of a parton pair at the short distance followed by the non-perturbative 
transition of the parton pair into the hadron pair when the virtuality of 
the virtual photon $Q^2$ is large enough and the invariant mass of 
the hadron pair $W^2$ is not too large \cite{muller-1994,diehl-2000}. 
The latter part can be parameterized in terms of the GDAs, which is defined
in analogy of of the LCDAs for a single hadron production. 
We now consider only at the leading order of $\alpha_s$, so that 
there appears only the quark GDAs:  
\begin{eqnarray}
\Phi_q^{h\bar{h}}(z,\zeta,W^2)=\int\frac{dx^-}{2\pi}e^{izP^+x^-}
\left.\langle h(p)\bar{h}(p^\prime)|\bar{q}(x)\gamma^+q(0)|0\rangle
\right|_{x^+=\vec{x}_\perp=0} ,
\label{quark-GDA}
\end{eqnarray} 
where $P=(p+p^\prime)/2$, $W^2=(p+p^\prime)^2$ and $\zeta=p^+/P^+$ 
is the momentum fraction of a hadron $h$ in the final $h\bar{h}$ pair. 
Also, the variable $z$ has a meaning of a momentum fraction carried 
by a quark in the intermediate quark pair. 
Although this process is obtained by the $s$-$t$ crossing of 
the deeply virtual Compton scattering (DVCS),  
it is not straightforward to relate the GDAs with the corresponding 
non-perturbative functions, generalized parton distributions (GPDs), 
which appear in DVCS \cite{gpd-gda-summary}. 
This is because the GDAs in the physical region 
corresponds to the GPD in the unphysical region \cite{ks-2013}.
In this work, we take a simple model for the GDAs 
($h=f_0(980)$ or $a_0(980)$) as 
\begin{eqnarray}
\Phi_q^{h\bar{h}}(z,\zeta,W^2)=N_{h(q)}z^\alpha(1-z)^\beta(2z-1)
\zeta(1-\zeta)F_{h(q)}(W^2) ,
\label{quark-GDA-2}
\end{eqnarray} 
which is consistent with the symmetry relations: 
$\Phi_q^{h\bar{h}}(1-z,1-\zeta,W^2)=\Phi_q^{h\bar{h}}(1-z,\zeta,W^2)
=-\Phi_q^{h\bar{h}}(z,\zeta,W^2)$ for the charge-conjugation  
even part of the GDAs \cite{diehl-2000}. Here $F_{h(q)}(W^2)$ is 
the quark form factor of the energy-momentum tensor which is related to 
the GDAs by a momentum sum rule \cite{gda-sum-rule}. 
Now we fix the overall factor $N_{h(q)}$
such that the sum of the second moments in 
$z$ amounts to 0.5 as the usual nucleon PDFs and parameterize 
the form factor as \cite{BC-1976}
\begin{eqnarray}
F_{h(q)}(W^2)=\frac{1}{[1+(W^2-4m_h^2)/\Lambda^2]^{n-1}} ,
\label{form-factor}
\end{eqnarray} 
where $\Lambda$ is the cutoff parameter and $n$ is the number of 
the constituents of the hadron $h$. Note that the complex phase 
from the final state interaction, etc., needs not be included 
because the bremmsstlahrung process can be neglected in this case, and  
only a single amplitude contributes to the cross section \cite{sk-2013}. 
\begin{figure}
\includegraphics[width=.45\textwidth]{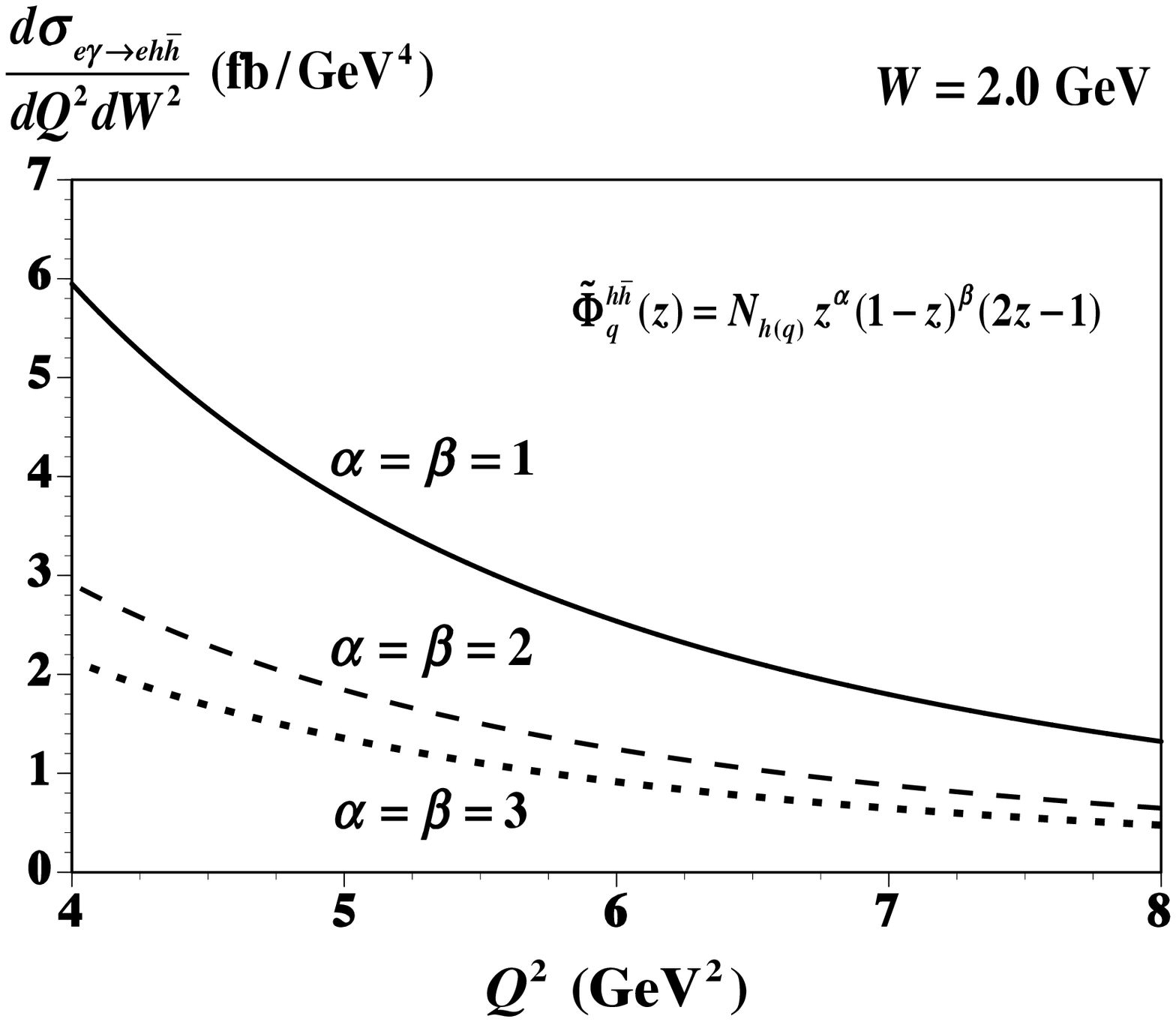}
\hspace{1cm}\includegraphics[width=.5\textwidth]{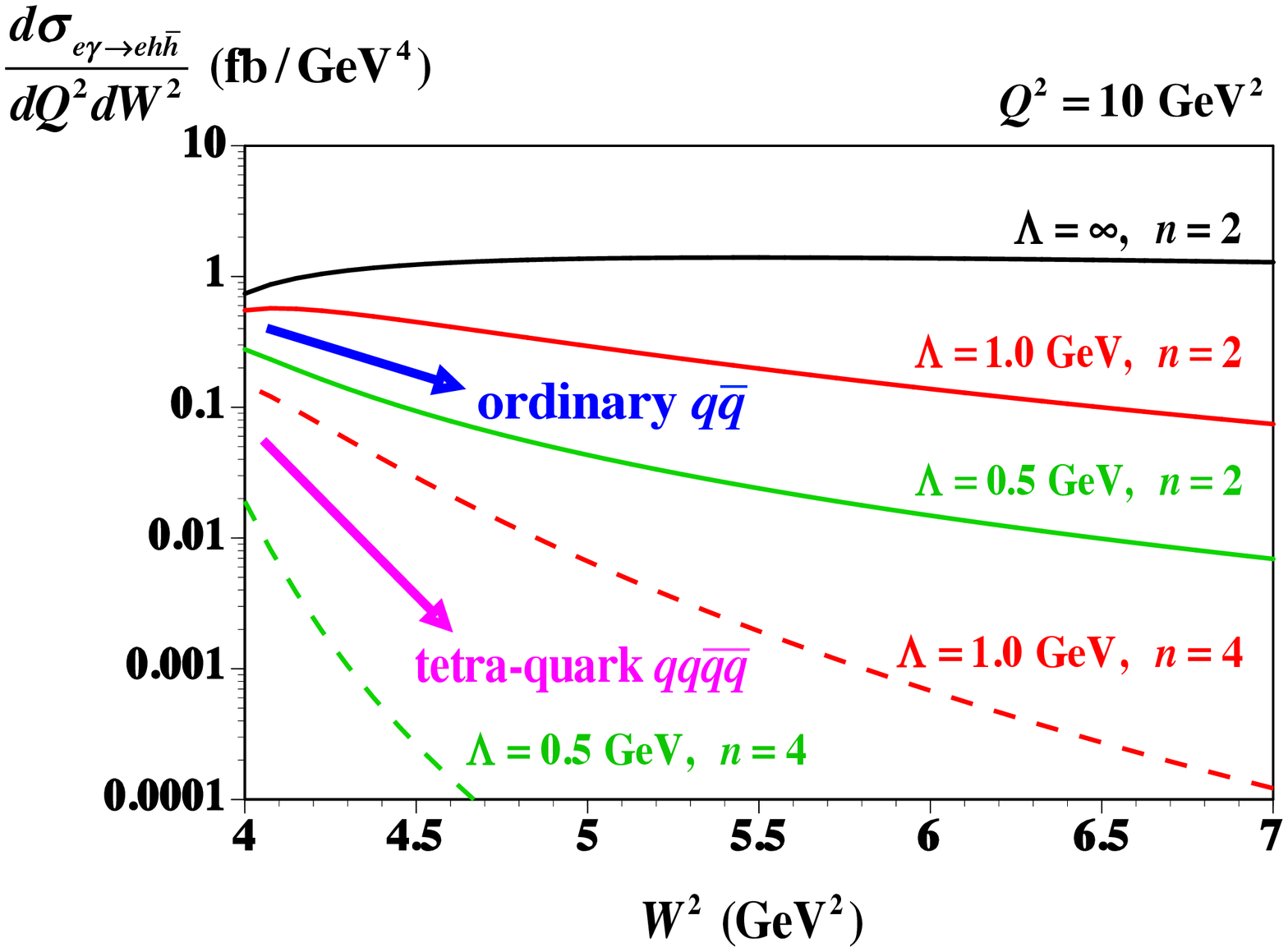}
\caption{Cross sections of $e+\gamma\rightarrow e+h+\bar{h}$ at $W=2$\,GeV 
as a function of $Q^2$ (left) and those at $Q^2=10$\,GeV$^2$ as a function of 
$W^2$ (right).}
\label{gda-cross}
\end{figure}

The left panel of Figure \ref{gda-cross} shows the differential 
cross sections, $d\sigma/dQ^2dW^2$, for the process 
$e+\gamma\rightarrow e+h+\bar{h}$ \cite{ks-2013} as a function of $Q^2$ 
at a fixed value of $W=2.0$\,GeV. Each line is calculated using the GDAs  
of eq.(\ref{quark-GDA-2}) with $(\alpha,\beta)=(1,1),(2,2),(3,3)$ and 
$\Lambda=1.0$\,GeV in eq.(\ref{form-factor}). 
The result is quite sensitive to the small-$z$ and large-$z$ behavior of 
the GDAs. In the right panel of Figure \ref{gda-cross}, we also show 
the same cross sections as a function of $W^2$ at a fixed $Q^2=10$\,GeV$^2$ 
with various values of the cutoff $\Lambda$ and $n$.  
Similarly to the results in the previous section, the cross sections 
assuming the ordinary $q\bar{q}$ state and those with the tetra-quark 
$qq\bar{q}\bar{q}$ state show quite a different behavior 
as a function of $W^2$, so that one can say that these observables
could be used to explore the internal structure of the scalar mesons.

\end{document}